\documentclass[prb,showpacs,twocolumn,superscriptaddress,amsmath,amssymb,notitlepage,aps]{revtex4}
\usepackage{graphicx}
\usepackage{dcolumn}
\usepackage{amsmath,bm}
\usepackage[mathlines]{lineno}
\usepackage{color}
\usepackage{pifont}

\usepackage[colorlinks=true,linkcolor=blue,citecolor=blue]{hyperref}
\usepackage[normalem]{ulem}
\allowdisplaybreaks[4]

\usepackage{txfonts}
\usepackage{soul,xcolor}
\usepackage{ulem}

\usepackage{xcolor,cancel}

\usepackage{soul,xcolor}
\setstcolor{red}

\begin{document}

\title{\bf Phase transition from Weyl to self-linked semimetal using bi-circular laser }

\author{Debashree  Chowdhury\footnote{Corresponding Author}}
\email{debashreephys@gmail.com}
\affiliation{Centre for Nanotechnology, IIT Roorkee, Roorkee, Uttarakhand  247667, India }
\date{\today}
\begin{abstract}
The Fermi surface topology of a triple non-hermitian (NH) Weyl-semimetal (WSM) driven by bi-circularly (BCL) polarized light is presented in this study. A NH WSM in particular has remarkable outlines. BCL light, however, modifies the symmetry features of NH triple Weyl and causes an unusual new kind of band swapping. We observe swapping between the imaginary bands (with or without exceptional degenaracies), which causes unique Fermi surfaces in the form of double rings and knots . We also discuss the corresponding changes in the Berry curvature as well.
\end{abstract}

\maketitle

\section{Introduction}
The search for new and distinct topological phases has long been at the forefront of research. Examples include the topological insulator (TI) \cite{Hasan, Kane}, nodal line semimetals, Dirac and Weyl-semimetal phases (WSM) \cite{Vafek, Burkov, Armitage}, and so on. The TIs are the first instance of the materials displaying unique edge states \cite{Kane}, even though the majority of the sample is insulating. The other semimetallic phases mentioned here provide a three-dimensional example \cite{Burkov, Armitage}. Different properties of the 3D WSMs have been remarkably studied in the literature \cite{Vafek, Burkov, Armitage}. 

The Weyl point, the source or sink of the Berry curvature, is the place where the two bands in WSM intersect. In this instance, the dispersion is linear, with a topological charge of $\pm 1$. Conversely, higher topological charge WSMs (i.e., $\pm 2$ or $\pm 3$) are found in nature. These are the multi-WSM (double or triple WSM) \cite{Gupta, MW2, MW1, TW, Fang}. The rotational symmetry of the point group $C_{n}$ is preserved in multi-weyl having a nonlinear dispersion \cite{Our 2023, Our,Huang}. WSMs with a $\pm 1$ charge are extensively studied\cite{Zhong,Armitage,NN,Scanlon}, but multi-WSMs have recently gained interest. The scope to find new topological phases in this unique class of materials is thus still active. Its study must illustrate the various characteristics of hermitian and NH multi-WSMs, further driven by time-periodic fields. 
\begin{figure}\label{Fig0}
	\includegraphics[width=0.45\textwidth]{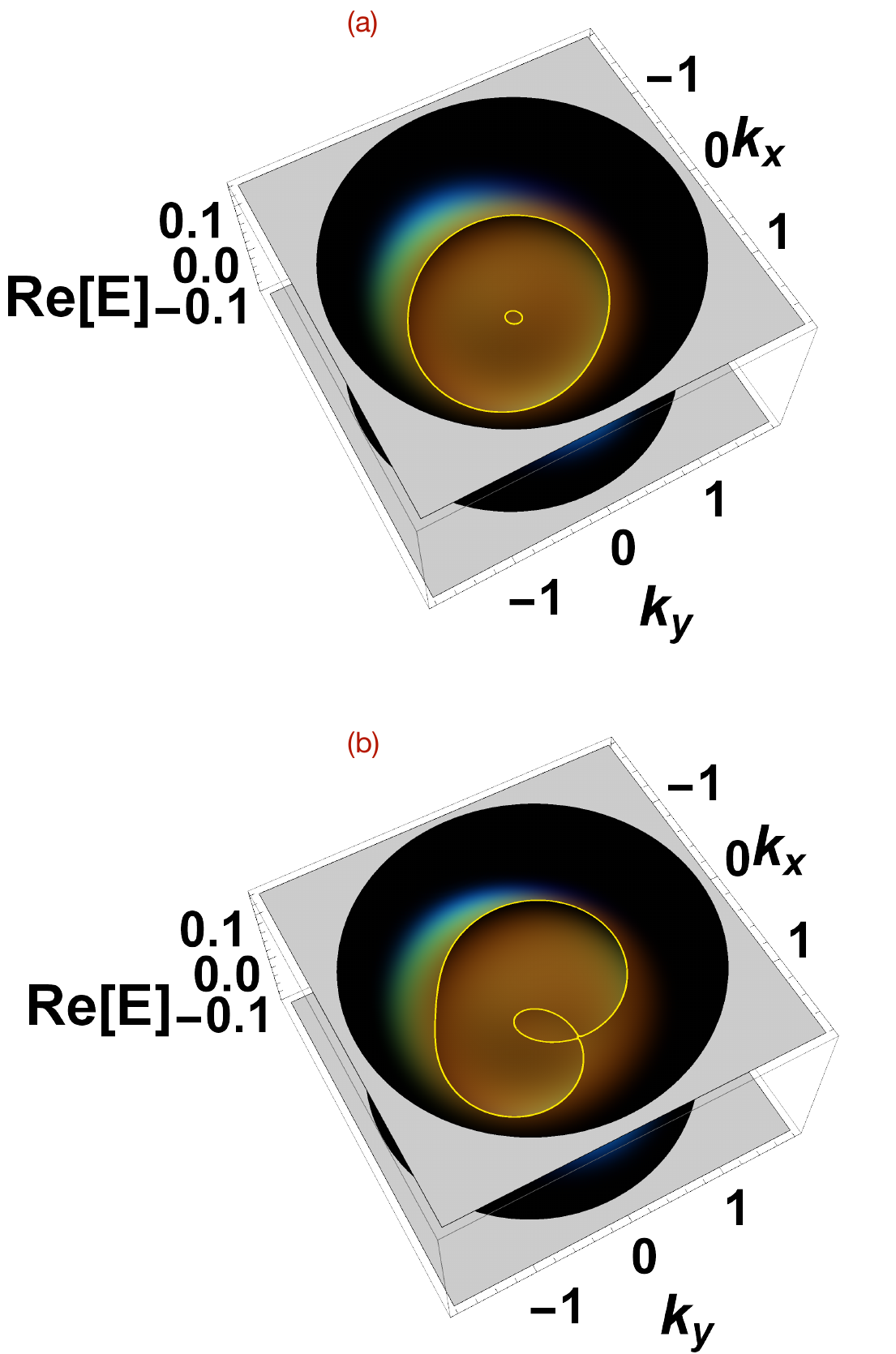}
	\caption{Schematic plot for the light-induced Lifshitz transitions in a triple WSM in the presence of a combined action of NH terms and BCL polarized light. The resulting Fermi surface shows a double nodal ring (a) and self-linked structures (b).}
\end{figure}
\begin{figure}\label{Fig1a}
	\includegraphics[width=0.48\textwidth]{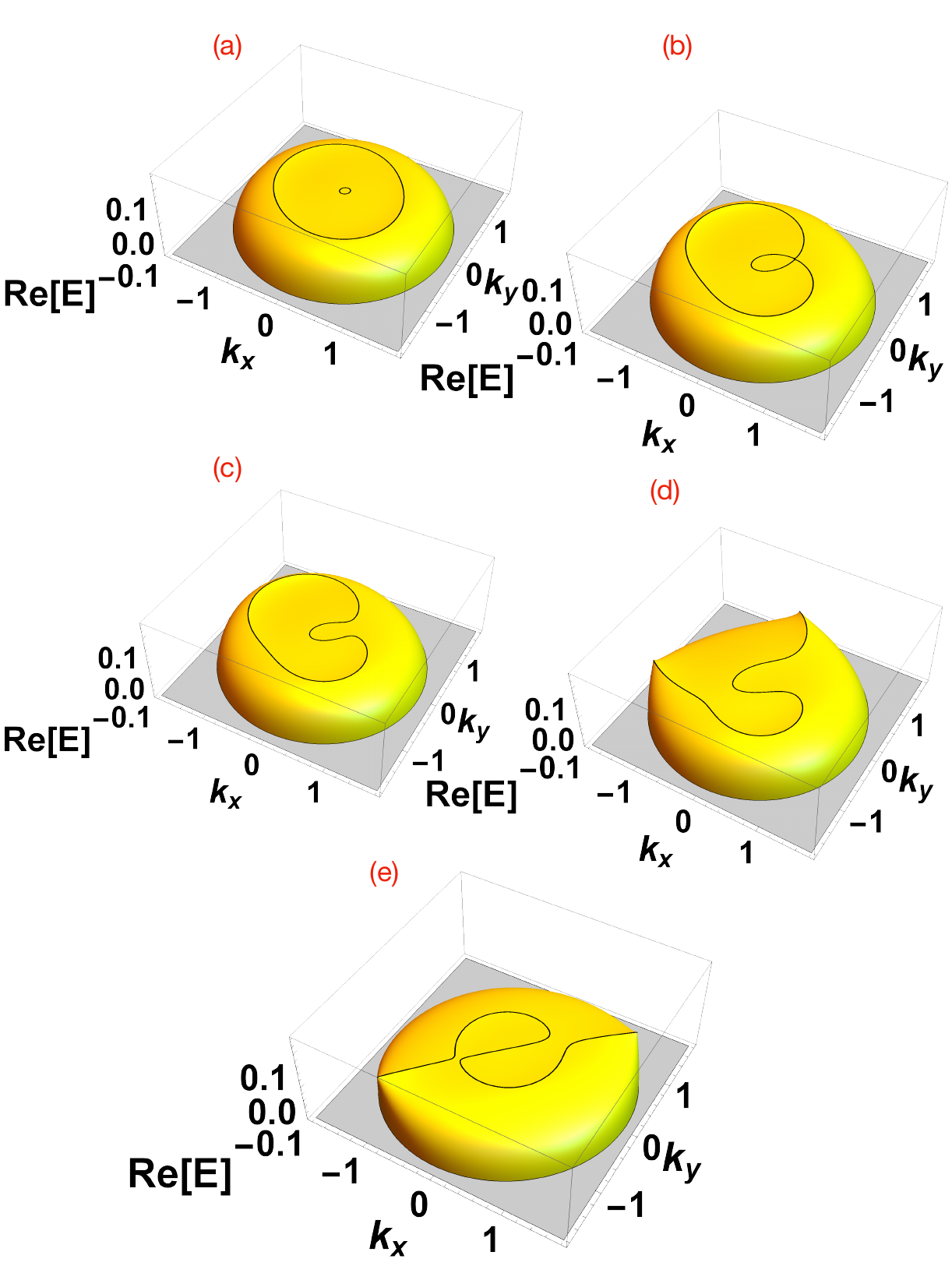}
	\caption{The energy vs $k_{x}$ and $k_{y}$ for different values of the angles $\beta,$ keeping $\alpha$ and  $\phi$ constant. In (a), the angle $\beta$ is considered to be $0.1$. The other angles are zero. In Fig. (b) for a slightly larger value of $\beta=0.262$ the double ring structure deviates into a self-linked structure. However, increasing the $\beta$ further ($\beta=0.3$), one gets the un-linked structure as indicated in Fig. (c). For further increase of $\beta $ produces open-ended curves indicated in figs (d), (e), for $\beta=0.5$ and $\beta=0.8$ respectively.}
\end{figure}
The light-induced topological phases greatly aid the comprehension of topological phase transitions \cite{f1,f2,f4,f5,f8,f13}. The Floquet mechanism is observed in topological materials when time-periodic fields are present. Previous research addresses light-induced topological phases when both linear and circular polarisations are present \cite{Our 2023, Our}. Remarkably, the frequency and amplitude of the light serve as the tuning parameters for the various phases. Furthermore, the distinct Fermi surface topology—which we refer to as Lifshitz transitions—is another feature of the light-induced topological phases \cite{f13a, f13c, Vargiamidis}. Though Floquet topological systems with circular or linear polarizations are widely addressed in the literature \cite{Vargiamidis,Tahir,Zhou}, bi-circular (BCL) polarization-induced Floquet dynamics have just been covered in \cite{bicircular, graphene}. "BCL" light refers to two superimposed circularly polarised lights with a frequency ratio of $\eta$ merging and having a phase difference $\alpha.$ The main idea of using the BCL light is to have increased tunability on different topological phases.  Importantly, BCL consists of two circularly polarized lights having different frequencies and has a crucial role in the spatial inversion symmetry and rotational symmetry of the system \cite{Ikeda}. The BCL shows a rose like pattern and has a rotational symmetry of ($n_{1}+n_{2})$/${\rm gcd(n_{1},n_{2})}$ fold, where $n_{1}$ and $n_{2}$ are the two frequencies of the circularly polarized lights that produce BCL. 


On the other hand, the NH perturbation in the Hamiltonian of the WSM is interesting as long as new and fascinating topological phases are concerned \cite{Luis1, Luis,Hamanaka}. For the details of NH topology, the reader can go through the review articles \cite{Bergholtz,Okuma}. However, the interesting issue, which makes the NH systems rich is the appearance of band swapping  \cite{Bergholtz,Okuma} and as a result, the exceptional points (EP) and exceptional contours (EC) \cite{ Ghatak, Gong, Bergholtz, Cerjan, Cerjan1}. We summarise here a few key details about the EPs, as this will help us discuss the results. The EPs are certain special degeneracy points, where not only the eigenvalues, but eigenvectors do also merge. One can show that this happens at certain positions of the band dispersion, where the real and imaginary bands touch each other. The NH double WSM are discussed in our previous papers \cite{Our 2023, Our}. In \cite{Our, Banerjee2}, it is shown that new topological phases arise in the presence of circularly polarised light. In the NH double WSM \cite{Our}, a captivating new concept of having new ECs from the parent contour is put forward. Also, the charge of the parent contour is divided into the two new ECs, with equal charges. In another paper \cite{Our 2023}, we have shown a similar system when the NH parameter has components in all three directions. This confirms that an NH hexagonally wrapped Fermi surface \cite{Fu, Basak} appears in the double WSM. 

\begin{figure}\label{Fig2}
	\includegraphics[width=0.48\textwidth]{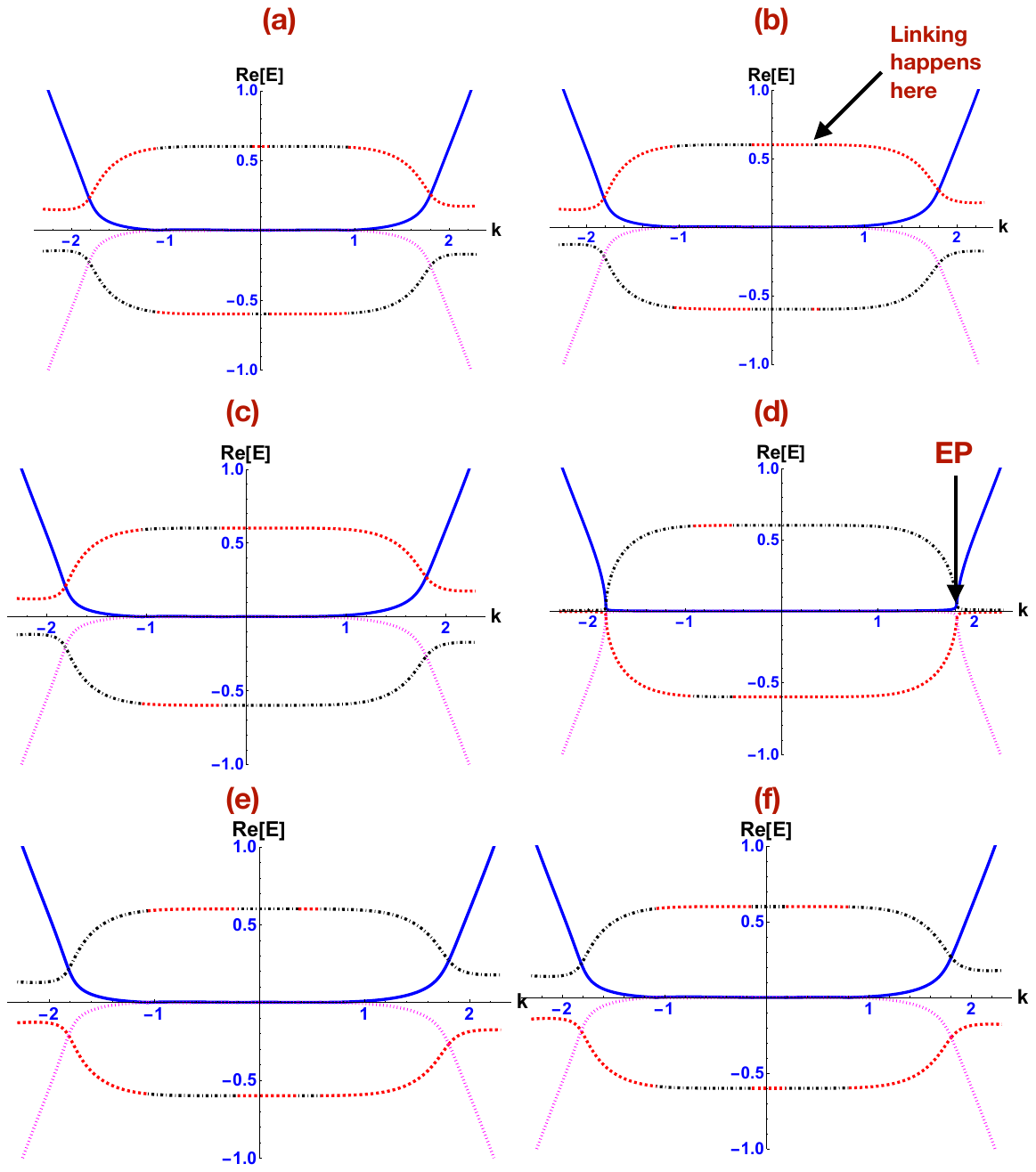}
	\caption{2D plot for energy vs $k=\sqrt{k_{x}^{2}+k_{y}^{2} }.$ In all the sub-figures, the real bands are indicated by the blue and magenta lines and the imaginary bands are presented by the red and black lines.  In (a) for $\beta=0.1,$ we have a mixing between imaginary bands (red and black lines). However, the real (blue and magenta lines) and imaginary bands touch at four places; these are degenerate points but do not show any exceptional behaviour. In (b), the imaginary band mixing happens in one place instead of two, as is the case in plot (a). Here we choose $\beta=0.262.$ Plot (c) has $\beta=0.3.$ Increasing values of $\beta=0.8$ in (d), one observes an interchange in the position of the imaginary bands. The whole story repeats for larger values of $\beta$ in (e) and (f). But the role of the imaginary bands is reversed.  In (b) and (d), we place black arrows to show the exact position where linking or exceptional points appear.}
\end{figure}

The present paper considers a BCL-polarised light-induced NH triple WSM, and the energy dispersion shows a unique Fermi surface topology. Fig. 1 shows a glimpse of the discussed Fermi surfaces (double nodal ring (Fig. 1(a)) and self-linked (Fig. 1(b))) in the paper. The novelty of our work lies in the fact that the topology of a NH triple WSM is tuned by the application of BCL light. The topological nature of Weyl systems in the presence of non-hermiticity has been discussed previously in the literature \cite{Our 2023, Our}. However, the topological phases we find in this work are hard to achieve with periodic circularly/linearly polarized light, even in the presence of non-hermiticity (this fact is discussed in appendix B for single WSM in presence of BCL and a triple WSM with a circular polarized light). The tuning of various topological NH phases with the different angles of the BCL light is something new and stands out from the line of work in NH topology. The sequence followed in the paper is as follows. In sec. II the model Hamiltonian is discussed in the presence of the BCL light. This section includes our main results regarding the appearance of the different nodal structures. Next, in sec. III, we compute the Berry curvature of the system and discuss the change in the nature of the curvature due to the change in the system parameters. We conclude in Section IV.

\section{The model Hamiltonian}
Let us start with a triple-WSM. The system Hamiltonian is given by \cite{Gupta, TW}
\begin{align}\label{triple}
H(k)=v_{z}k_{z}\sigma_{z}+ a\Big(k^{3}_{-}\sigma_{-}+k^{3}_{+}\sigma_{+}\Big),
\end{align}
where we denote $k_{\pm}=k_{x}+ik_{y}$ and $\sigma_{\pm}=\frac{1}{2}\Big(\sigma_{x}+i\sigma_{y}\Big).$ Here $v_{z}$ is the Fermi velocity and $a=\frac{1}{2m}.$ The energy eigenvalues for this system are given by
\begin{align}
{\cal E}_{\pm}=\pm \sqrt{v^{2}_{z}k_{z}^{2}+a^{2}\Big(k_{x}^{2}+k_{y}^{2}\Big)_{}^{3}}.
\end{align}

As the next step, we would like to investigate the role of BCL polarized light on the triple WSM. BCL appears due to the superposition of two circularly polarized lights having two opposite chiralities and different frequencies of an integer multiple of $\eta$ and maintaining a phase difference $\alpha.$ The polarization field induced by a BCL light is \cite{bicircular}
\begin{align} 
A(t)=A_0\sqrt{2}\,\text{Re}\left[e^{-i(\eta\Omega t-\alpha)}\bm{\epsilon}_{R}+e^{-i\omega t}\bm{\epsilon}_{L}\right]\,,
\label{2}
\end{align}
where $A_0$ is the amplitude and $\bm{\epsilon}_{L/R}$ are left (L) and right (R) circularly polarized (CL) basis vectors. 
A more generic incidence direction can be constructed by rotating the incident plane by an angle $\beta$ around $\hat{z}$. Following \cite{bicircular}, we write the BCL polarized  vector potential as

\begin{widetext}
\begin{align}\label{BCL}
A_{x}(t)&=A_{0}\Big[\cos\beta\Big(\cos(\eta\Omega t-\alpha)+\cos\Omega t\Big)+\sin\beta \cos\phi \Big(\sin\Omega t-\sin(\eta\Omega t-\alpha)\Big)\Big],\nonumber\\
A_{y}(t)&=A_{0}\Big[\sin\beta\Big(\cos(\eta\Omega t-\alpha)+\cos\Omega t\Big)+\cos\beta \cos\phi \Big(\sin\Omega t-\sin(\eta\Omega t-\alpha)\Big)\Big],\nonumber\\
A_{z}(t)&=-A_{0} \sin\phi \Big(\sin\Omega t-\sin(\eta\Omega t-\alpha)\Big),
\end{align}
\end{widetext}
where $\phi$ denotes the incident angle. 
In the presence of the gauge fields, the momentum in three directions changes as $k_{i}\rightarrow k_{i}+\frac{e A_{i}}{\hbar}.$ Thus we rewrite the Hamiltonian in presence of the BCL light \cite{Our}
\begin{align}
H_{1}=H(k)+V(t),
\end{align} 
where $V(t)$ is the time-dependent part of the Hamiltonian.  We here use the high frequency approximation, where it is considered that the applied frequency is much higher than any other energies of the system (say for example the band gap). 
Using Floquet formalism in the high-frequency limit, the effective Hamiltonian is obtained as
\begin{align}\label{6}
H_{eff}=H(k)+\frac{1}{\hbar \Omega}\Big[V_{-1},V_{+1}\Big],
\end{align} 
where $V_{\pm}=\int_{0}^{T} dt e^{\pm i\Omega t}V(t).$ Here $T=\frac{2\pi}{\Omega}$ is the time period. This is to be noted that although $\eta$ can take any integer value, we consider $\eta=2.$ The effective Hamiltonian for a triple WSM (n=3) in the presence of BCL polarized light can be obtained from the high frequency approximation stated in Eq. (\ref{6}). However, we are interested here in the NH properties of such a BCL polarized triple WSM. In search of this we by hand add a loss/gain term $i\gamma\sigma_{z},$ where $\gamma$ is the parameter for non-hermiticity. Thus we finally have the effective Hamiltonian with the NH loss gain term as follows 

\begin{align}\label{Heff}
H_{eff}&=\Big(v_{z}k_{z}+i\gamma+\frac{i}{\hbar \Omega}({\cal N}_{1}{\cal N}_{4}-{\cal N}_{2}{\cal N}_{3})\Big)\sigma_{z}\nonumber\\&+\Big(a\Big(k_{y}^{3}-3k_{x}^{2}k_{y}\Big)+\frac{i}{\hbar \Omega}({\cal N}_{1}+{\cal N}_{3}){\cal N}_{5}\Big)\sigma_{y}\nonumber\\&+\Big(a\Big(k_{x}^{3}-3k_{x}k_{y}^{2}\Big)-\frac{i}{\hbar \Omega}({\cal N}_{2}+{\cal N}_{4}){\cal N}_{5}\Big)\sigma_{x}
,
\end{align}
where ${\cal N}_{i}$ are the function of $k_{x},k_{y},\phi,\alpha,\beta$ and $A_{0}.$ The exact expressions are quite cumbersome and are written in Appendix A.

Interestingly, in Eq. (\ref{Heff}), $i\gamma\sigma_{z}$ is the NH perturbation, which is added here by hand to understand the NH nature of the triple WSM in the presence of BCL polarised light. In the later part of the paper, we discuss situations where another loss/gain term exists in the $\sigma_{x}$ direction. In experiments, those terms can be incorporated by using a coupled-resonator optical waveguide having lopsided internal scattering \cite{Marques}. These loss/gain terms can be achieved by spatially engineered gain and loss \cite{Marques} or in finite lifetime quasiparticles. Due to electron-electron, electron-phonon scattering, the electron's self energy brings the non-hermiticity in the system. In \cite{Fu}, the authors has shown that the finite lifetime of the quasiparticles adds a gain/loss coupling term along $\sigma_{z}$. Experimentally, the $\sigma^z$ term in a hermitian system corresponds to asymmetry in the internal modes or within two sublattices. However, an imaginary  term associated to the $\sigma_{z}$ indicates that one of the two sublattices experiences gain whereas the other sublattice goes through loss. The situation might also be like: both sublattices experience loss/gain of different degrees. This creates a NH diagonal term (along $\sigma_{z}$). In \cite{Cerjan}, the authors have elaborated the process of having the gain along $\sigma_{x}$ in a photonic woodpile structure. In \cite{Cerjan} the authors discuss a layered chiral woodpile photonic crystal, that operates in the THz regime. The unit cell consists of A and B sublattices, which are coupled via an asymmetric junction. This unique structure generates a complex off-diagonal hopping term: the $\sigma_x$ couplings.
The energy eigenvalues of Eq. (\ref{Heff}) are written as
\begin{align}
&{\cal E}_{\pm}=\pm\frac{1}{\Omega }\Big(F(\beta,\gamma,\phi,\alpha)\Big)^{1/2},
\end{align}
where $F(\beta,\gamma,\phi,\alpha)$ is a function of all the parameters of the theory. The final expression of $F(\beta,\gamma,\phi,\alpha)$ is presented in the appendix A.
Interestingly the results can be tuned by regulating the angles associated with the BCL polarized light. We add different figures supporting different Fermi surface topologies. Let us start with the combination where all three angles are zero and we have a circular Fermi surface in the real part of the energy. We fix the two angles to zero while tuning $\beta$. When $\beta=0.1,$ one reaches the Fermi surface of a double ring structure as is indicated in Fig. 2(a). 
Increasing $\beta$ further, at a value $\beta=0.262,$ the double ring structure deviates and starts producing a nodal knot (see Fig. 2(b)). However, for larger values of $\beta$ the nodal knot produces an un-linked closed structure (see Fig. 2(c) for $\beta=0.3$). If we keep on increasing $\beta=0.5,$ the un-linked closed loop deformed into an open-ended arc, which is shown in Fig. 2(d). The arc is reshaped for $\beta=0.8$ (Fig. 2(e)). For $\beta=1.1,$ we recover the un-linked structure. For $\beta=1.3$ the self-linked structure is regained. Further increasing $\beta$ returns back double nodal structure at $\beta=1.35$. An obvious question appears what happens if we keep increasing $\beta$ further?  This repeats the whole story. Thus one may conclude that the system has a periodicity in $\beta.$ 

However, the above situation may also appear if we choose to vary $\phi,$ keeping the other two angles at zero or properly choosing values for all the angles.

\begin{figure}\label{Fig4}
\includegraphics[width=0.48\textwidth]{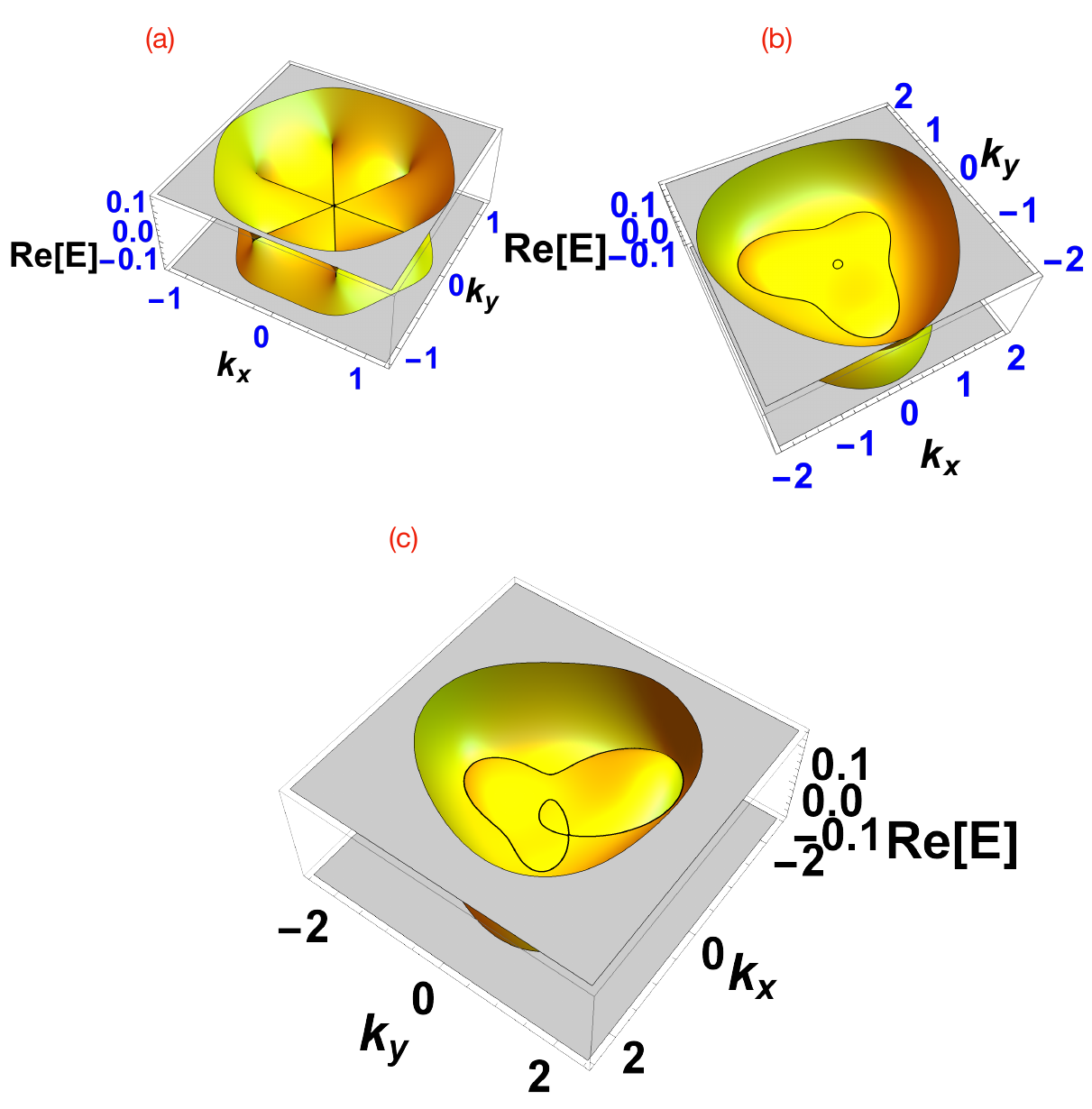}
\caption{3D plot for energy vs k. In Fig. (a) we have a hexagonally symmetric pattern for NH parameter along $\sigma_{z},$ as zero and along $\sigma_{x}$ is 0.1. For $\beta=0.1,$ keeping other angles as zero and for NH parameter along $\sigma_{x}$ is 0.09  and NH parameter along $\sigma_{z}$ is 0.9, we achieve a double nodal structure. Changing $\beta$ to 0.263, we achieve a self-linked structure in this case as well.}
\end{figure}

It is important to state here that the different structures, we obtain by changing the angles of the BCL polarized light, has also a crucial dependence on the NH parameter $\gamma.$ It then demands a discussion on whether the obtained structures are exceptional. To discuss this, we present here the 2D diagrams of ${\cal E}_{\pm}$ vs. $k$ in Fig. (3). Here $k_{x}=k\cos({\cal M})$ and $k_{y}=k\sin({\cal M}),$ and $k=\sqrt{k_{x}^{2}+k_{y}^{2}}.$ In Fig.3 (a), it is shown that when the other two angles are kept to be zero and $\beta=0.1$, the real and imaginary bands touch each other at four points, but the meeting does not take place at zero energy. Thus although the points are degenerate but not exceptional ones. However, the imaginary energy bands swap between each other at two sides of the zero energy and we have a double-ring structure (Fig. 3(a)). For $\beta=0.262,$ although we have the same number of degenerate points, where the real and imaginary bands meet, the band swapping happens in only one place between the imaginary bands, which is indicated in Fig. 3(b). As a result, we obtain the self-linked structure. For $\beta=0.3,$ in Fig. 3(c), the 2D band diagram of un-linked closed loop is shown.  Further increase of $\beta=0.8,$ provides only two places, where the real and imaginary bands meet at zero energy, which is one of the conditions to have EPs. For $\beta=0.8,$ we have two exceptional points. However, in this case, the imaginary bands change their positions completely and we have an open-ended arc as indicated in Fig. 3 (d). If we further increase the angle $\beta,$ surprisingly we get the self-linked structure back at $\beta=1.3$ (Fig. 3(e)). As in the previous case, the number of degenerate points is four, but the position of the imaginary bands is switched. Similarly for $\beta=1.35$, in Fig. 3(f) the double nodal structure, with switched imaginary bands is obtained. This shows multiple band swapping between the imaginary bands. 

The above discussion considers the NH parameter along the $\sigma_{z}$ direction. However, one may also have a situation where the non-hermiticity arises in both directions of $\sigma_{x}$ and $\sigma_{z}.$ In the first case, when we keep NH parameter along $\sigma_{z},$ to be zero and along $\sigma_{x}$ is 0.1, we get a hexagonally symmetric Fermi surface for any value of $\beta,$ with $\alpha,~\phi=0,$ which is presented in Fig. 4 (a). However, adding the non-hermiticity along $\sigma_{z}$ as well, one can observe changes in the previous situation. For $\beta=0.1,$ keeping other angles as zero and for NH parameter along $\sigma_{x}$ is 0.09  and NH parameter along $\sigma_{z}$ is 0.9, we achieve a double nodal structure (see Fig. 4(b)), but the rings are not circular. Although the small one is circular, however the big one is curvy. For particular values of the NH parameters in two directions and $\beta=0.263$, one achieves a self-linked structure here as well (see Fig. 4 (c)).
\begin{figure}\label{Fig1}
\includegraphics[width=0.4\textwidth]{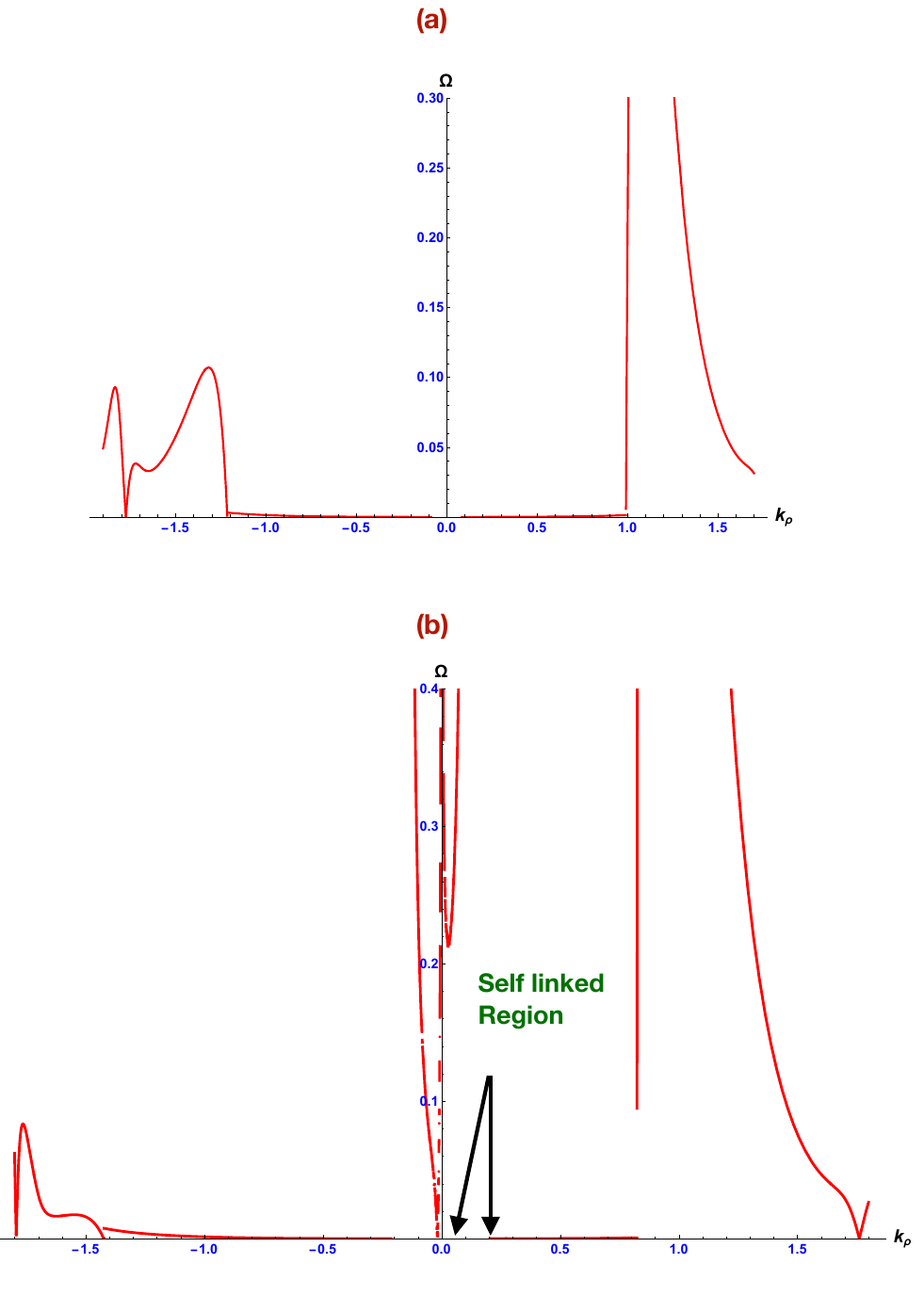}
\caption{Berry curvature vs $k_{\rho}$ plot. In (a), $\beta$ is chosen to be 0.1, whereas other angles are chosen to be zero. This provides a discontinuity in the plot at two places around $k_{\rho}=0$ ($k_{\rho}<0.2$) and $k_{\rho}\approx 0.85.$  In (b) ($\beta=0.262$) the Berry curvature shows a discontinuity around zero ($k_{\rho}<0.08$) but with a central part around zero and also shows discontinuities at $k_{\rho}= 0.8.$  The little central part in Fig. 5 (b) shows the place where linking happens (indicated by two arrows) in the self-linked structure.} 
\end{figure}
\section{The Berry curvature}
\begin{figure}\label{Fig6}
\includegraphics[width=0.4\textwidth]{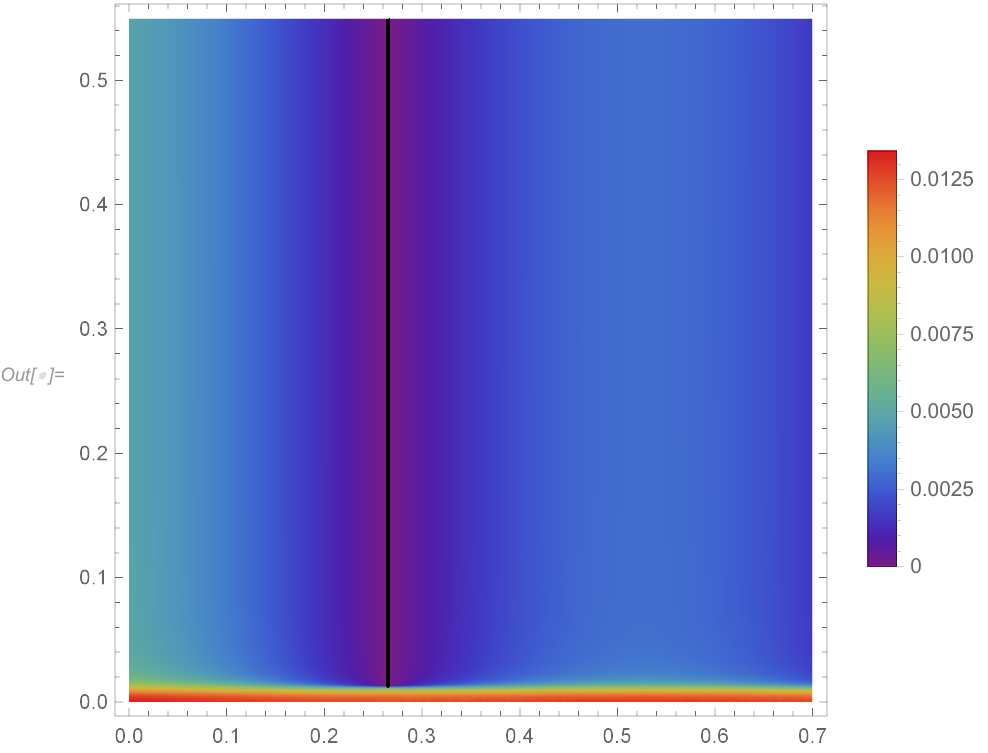}
\caption{Phase diagram for different values of $\gamma$ and $\beta$, for $k_{x}=0.3,$ $k_{y}=0$ and $k_{z}=0.$ The value for $A_{0}=0.5.$} 
\end{figure}
We are now in the position to discuss the topological nature of the two important structures- the double nodal ring and the self-linked structure. This can be done using the Berry curvature analysis. The left and right eigenstates of the effective Hamiltonian in Eq. (\ref{Heff}) are \cite{Cerjan,Cerjan1, Our}
\begin{align}\label{B1}
  \Big \langle \psi^{L}\Big|&= \Big[\Big(a\Big(k_{x}^{3}-3k_{x}k_{y}^{2}\Big)-\frac{i}{\hbar \Omega}({\cal N}_{2}+{\cal N}_{4}){\cal N}_{5}\Big)\nonumber\\&+i\Big(a\Big(k_{y}^{3}-3k_{x}^{2}k_{y}\Big)+\frac{i}{\hbar \Omega}({\cal N}_{1}+{\cal N}_{3}){\cal N}_{5}\Big),\nonumber\\&{\cal E} -i \gamma-\Big(v_{z}k_{z}+\frac{i}{\hbar \Omega}({\cal N}_{1}{\cal N}_{4}-{\cal N}_{2}{\cal N}_{3})\Big) \Big]\nonumber\\
\Big | \psi^{R}\Big\rangle&=  \Big[\Big(a\Big(k_{x}^{3}-3k_{x}k_{y}^{2}\Big)-\frac{i}{\hbar \Omega}({\cal N}_{2}+{\cal N}_{4}){\cal N}_{5}\Big)\nonumber\\&-i\Big(a\Big(k_{y}^{3}-3k_{x}^{2}k_{y}\Big)+\frac{i}{\hbar \Omega}({\cal N}_{1}+{\cal N}_{3}){\cal N}_{5}\Big),\nonumber\\&{\cal E} -i \gamma-\Big(v_{z}k_{z}+\frac{i}{\hbar \Omega}({\cal N}_{1}{\cal N}_{4}-{\cal N}_{2}{\cal N}_{3})\Big)\Big]^{T}.
\end{align}
 We consider the cylindrical coordinates ($k_{\rho},\chi,k_{z}$) with $k_{x}=k_{\rho}\cos(\chi)$ and $k_{y}=k_{\rho} \sin(\chi).$ The Berry charge is written as,
 ${\mathfrak C}=\int_{C} {\bf\Omega}^{LR}({\textbf k})\cdot d{\bf S},$
	where ${\bf\Omega}^{LR}({\textbf k})=\nabla \times  {\cal A}^{LR}({\textbf k})$ is the Berry curvature. The superscripts on the Berry curvature denote that it is derived using the left and right eigenfunctions of the NH Hamiltonian and ${\cal A}^{LR}({\textbf k})$ is the Berry gauge field, which is obtained as
 \begin{align}
{\cal A}^{LR}(k)=i\left<\psi^{L}(k)|\nabla|\psi^{R}(k)\right>.
 \end{align}
 Next, we discuss the computed Berry curvature, which has three components ($\Omega_{k_{\rho}},$ $\Omega_{\chi},$ and $\Omega_{k_{z}}$) in cylindrical coordinates. The expressions of the three components of the Berry curvatures are cumbersome, and we do not present the explicit expressions here. However, we present plots for $\Omega =\sqrt{\Omega^{2}_{k_{\rho}}+\Omega^{2}_{\chi}+\Omega^{2}_{k_{z}}}$ taking $\alpha=\phi=0$ and varying $\beta.$ In Fig. 5, we have presented the results for the the two cases of double nodal ring and self-linked structures. In the first case, similar to the plots in Fig. 3 (a), when $\beta=0.1,$ we have swapping between the imaginary band at two places and this causes divergence in the Berry curvature at two different places. One is around $k_{\rho}=0,$ which causes the small circle of the two-ring structure. Another divergence is found near the $k_{\rho}=+0.85.$ This causes the larger circle in the double ring pattern. The situation is presented in Fig. 5 (a). In Fig. 5 (b), the Berry curvature for $\beta=0.262$ is plotted and the situation shows divergence around $k_{\rho}=0,$ with a small central part. The other divergences occur at $k_{\rho}=0.8.$ The central part in fig. 5 (b) causes the self-linked Fermi surface. Thus one concludes the discontinuity in the Berry curvature is related to the band swapping and Fermi surface topology. Here we would like to provide a phase diagram for fixed values of $k_{x}$ and $k_{y}.$ This is indicated in Fig. 6. The self-linking is formed for $\beta=0.262,$ and the phase diagram shows a minimum of energy at that value. This is shown by the violet region in the plot.
 
\section{Discussion and Conclusion}
In this paper, we have analysed a NH triple WSM in the presence of a BCL polarized light. This kind of polarization offers better tunability and provides modification in spatial inversion and rotational symmetry of the crystal. This combined action generates unique topological phases. The system we choose is the triple WSM, which has a $C_{6}$ rotational symmetry, having a non-linear dispersion. However, once the BCL light is switched on, the light modifies the rotational symmetry and in a combined effect of non-hermiticity and the BCL, we eventually land up in a double nodal ring and self-linked nodal line semimetals. This is the novelty of this work. The BCL light can be thought of as an important tool to switch between triple WSM to different nodal structures. We have also discussed the topological nature of the structures by analysing the Berry curvature. The important question in this regard is: can the combined action of the NH term and BCL light generate similar Fermi surfaces for charge 1 WSM? The answer is no. This is since the BCL significantly modifies the spatial inversion and rotational symmetry of the system. The multi-WSM, itself has a rotational symmetry of $C_{n},$ this when modified by BCL put forward interesting scenario.


\noindent \textit{Acknowledgments:} D.C. acknowledges financial support from DST (project number  DST/WISE-PDF/PM-40/2023). D.C would also like to thank Dr. Awadhesh Narayan for several discussions.
\begin{widetext}
\appendix
\section{The effective Hamiltonian and its components}
The effective Hamiltonian after the Floquet high frequency approximation in presence of a loss/gain term is written as
\begin{align}
H_{eff}&=v_{z}k_{z}\sigma_{z}+ a\Big(k^{3}_{-}\sigma_{-}+k^{3}_{+}\sigma_{+}\Big)+\frac{i}{\hbar \Omega}({\cal N}_{1}{\cal N}_{4}-{\cal N}_{2}{\cal N}_{3})\sigma_{z}+\frac{i}{\hbar \Omega}({\cal N}_{1}+{\cal N}_{3}){\cal N}_{5}\sigma_{y}\nonumber\\&-\frac{i}{\hbar \Omega}({\cal N}_{2}+{\cal N}_{4}){\cal N}_{5}\sigma_{x},
\end{align}
where
\begin{align}
   & {\cal N}_{1}=\frac{1}{32} A_{0} \Big(36 a A_{0}^2 \cos ^3(\beta )-27 a A_{0}^2 \cos (\beta )+27 a A_{0}^2 \cos (3 \beta )+2 \cos (\beta ) \Big(-3 a \left(3 A_{0}^2-8 k_{x}^2+8 k_{y}^2\right)\nonumber\\&+9 a A_{0} \left(A_{0}\sin ^2(\phi )+4 i k_{y} \sin (\alpha ) \sin (\beta )\right)-3 a A_{0} \cos ^2(\phi ) (3 A_{0}+8 i k_{y} \sin (\alpha ) \sin (\beta ))+48 i a k_{x} k_{y} \cos (\phi )\Big)\nonumber\\&+\frac{1}{2} i \Big(3 \sin (\beta ) \left(9 a A_{0}^2 \cos (2 \beta ) \cos (\phi )-6 a A_{0}^2 \sin ^2(\beta ) \cos (3 \phi )+15 a A_{0}^2 \cos (\phi )-32 a k_{x}^2\cos (\phi )+64 i a k_{x} k_{y}+32 a k_{y}^2 \cos (\phi )\right)\nonumber\\&+8 A_{0} \sin (\alpha ) (-3 a k_{x} \cos (2 \beta ) (\cos (2 \phi )+3)+\cos (2 \phi ) (3 a k_{x}-3 a k_{y} \sin (2 \beta ))+9 a k_{x}+24 i a k_{y} \cos (\phi ))\Big)\nonumber\\&+3 a A_{0} \cos ^2(\beta ) \left(36 i A_{0} \sin (\beta ) \cos ^3(\phi )-8 i k_{x} \sin (\alpha ) (\cos (2 \phi )+3)\right)\nonumber\\&+8 A_{0} \cos (\alpha ) \left(3 a k_{x} \cos (2 \beta ) (\cos (2 \phi )+3)-6 a k_{y} \sin (2 \beta ) \sin ^2(\phi )-12 i a k_{y} \cos (\phi )\right)\Big),\nonumber\\
{\cal N}_{2} &=\frac{1}{32} A_{0} \Big(-36 i a A_{0}^2 \cos ^3(\beta ) \cos ^3(\phi )+12 i \cos (\beta ) \Big(9 a A_{0}^2 \sin ^2(\beta ) \cos ^3(\phi )+3 a A_{0}^2 \cos (\phi )-4 a A_{0} k_{x} \sin (\alpha ) \sin (\beta ) \cos ^2(\phi )\nonumber\\&+6 a A_{0} k_{x} \sin (\alpha ) \sin (\beta )+4 a k_{x}^2 \cos (\phi )+8 i a k_{x} k_{y}-4 a k_{y}^2 \cos (\phi )\Big)+\frac{1}{2} i \Big(\sin (\beta ) \Big(36 i a A_{0}^2 \cos (2 \beta )+4 i \left(9 a A_{0}^2 \cos (2 \phi )+24 a k_{x}^2-24 a k_{y}^2\right)\nonumber\\&+192 a k_{x} k_{y} \cos (\phi )\Big)+8 A_{0} \sin (\alpha ) (-3 \cos (2 \phi ) (a k_{x} \sin (2 \beta )+a k_{y})+24 i a k_{x} \cos (\phi )+3 a k_{y} \cos (2 \beta ) (\cos (2 \phi )+3)-9 a k_{y})\Big)\nonumber\\&+8 A_{0} \cos (\alpha ) \left(-6 a k_{x} \sin (2 \beta ) \sin ^2(\phi )-12 i a k_{x} \cos (\phi )-3 a k_{y} \cos (2 \beta ) (\cos (2 \phi )+3)\right)\nonumber\\&+3 a A_{0} \cos ^2(\beta ) (-36 A_{0} \sin (\beta )+8 i k_{y} \sin (\alpha ) (\cos (2 \phi )+3))\Big),\nonumber\\
{\cal N}_{3} &=\frac{1}{32} A_{0} \Big(-9 i a A_{0}^2 \sin (3 \beta ) \cos ^3(\phi )+36 a A_{0}^2 \cos ^3(\beta )-27 a A_{0}^2 \cos (\beta )+27 a A_{0}^2 \cos (3 \beta )\nonumber\\&+\cos (\beta ) \Big(-18 a A_{0}^2 \cos ^2(\phi )-9 a A_{0}^2+3 a A_{0} \left(-3 A_{0} \cos (2 \phi )-32 i k_{y} \sin (\alpha ) \sin (\beta ) \sin ^2(\phi )\right)+48 a \left(k_{x}^2-k_{y}^2\right)-96 i a k_{x} k_{y} \cos (\phi )\Big)\nonumber\\&+\frac{1}{4} \sin (\beta ) \left(-i \cos (\phi ) \left(63 a A_{0}^2+192 a \left(k_{y}^2-k_{x}^2\right)\right)+27 i a A_{0}^2 \cos (3 \phi )-384 a k_{x} k_{y}\right)\nonumber\\&+3 a A_{0} \cos ^2(\beta ) \left(8 i k_{x} \sin (\alpha ) (\cos (2 \phi )+3)-36 i A_{0} \sin (\beta ) \cos ^3(\phi )\right)+12 i a A_{0} k_{x} \sin (\alpha ) \cos (2 \beta ) (\cos (2 \phi )+3)\nonumber\\&+8 A_{0} \cos (\alpha ) \Big(3 a k_{x} \cos (2 \beta ) (\cos (2 \phi )+3)-6 a k_{y} \sin (2 \beta ) \sin ^2(\phi )+12 i a k_{y} \cos (\phi )\Big)-12 i A_{0} \sin (\alpha ) (a k_{x} \cos (2 \phi )+3 a k_{x}-8 i a k_{y} \cos (\phi ))\Big),\nonumber\\
{\cal N}_{4} &= \frac{1}{32} A_{0} \Big(36 i a A_{0}^2 \cos ^3(\beta ) \cos ^3(\phi )-9 a A_{0}^2 \sin (3 \beta )+\cos (\beta ) \Big(-108 i a A_{0}^2 \sin ^2(\beta ) \cos ^3(\phi )-4 i \cos (\phi ) \left(9 a A_{0}^2+12 a k_{x}^2-12 a k_{y}^2\right)\nonumber\\&-96 i a A_{0} k_{x} \sin (\alpha ) \sin (\beta ) \sin ^2(\phi )-96 a k_{x} k_{y}\Big)+\sin (\beta ) \left(-18 a A_{0}^2 \cos (2 \phi )+9 a A_{0}^2-48 a k_{x}^2-96 i a k_{x} k_{y} \cos (\phi )+48 a k_{y}^2\right)+8 A_{0} \cos (\alpha ) \nonumber\\&\left(-6 a k_{x} \sin (2 \beta ) \sin ^2(\phi )+12 i a k_{x} \cos (\phi )-3 a k_{y} \cos (2 \beta ) (\cos (2 \phi )+3)\right)+12 i A_{0} \sin (\alpha ) (8 i a k_{x} \cos (\phi )+a k_{y} \cos (2 \phi )+3 a k_{y})\nonumber\\&+3 a A_{0} \cos ^2(\beta ) (-36 A_{0} \sin (\beta )-8 i k_{y} \sin (\alpha ) (\cos (2 \phi )+3))-12 i a A_{0} k_{y} \sin (\alpha ) \cos (2 \beta ) (\cos (2 \phi )+3)\Big),\nonumber\\
{\cal N}_{5}&=\frac{1}{2} i A_{0} v_{z} \sin (\phi ).
\end{align}
$N_{i}$'s are the function of system parameters. These $N_{i}$'s are determined for different values of the angles. 
The eigen-values are presented in terms of $F(\beta,\gamma,\phi,\alpha)$ with
\begin{align}
&F(\beta,\gamma,\phi,\alpha)=\Big(3 a^2 k_{x}^{4} k_{y}^2 \Omega ^2+3 a^2 k_{x}^2 k_{y}^4 \Omega ^2+a^2 k_{x}^6 \Omega ^2+a^2 k_{y}^6 \Omega ^2+2 N_{2} \left(-i \Omega  \left(a k_{x}^3  {\cal N}_{5}-3 a k_{x} k_{y}^2  {\cal N}_{5}+k_{z}  {\cal N}_{3} v_{z}+i \gamma   {\cal N}_{3}\right)+ {\cal N}_{1}  {\cal N}_{3}  {\cal N}_{4}- {\cal N}_{4}  {\cal N}_{5}^2\right)\nonumber\\&-2  {\cal N}_{1} \left(-i a k_{y}  {\cal N}_{5} \Omega  \left(k_{y}^2-3 k_{x}^2\right)+ {\cal N}_{4} \Omega  (\gamma -i k_{z} v_{z})+  {\cal N}_{3}  {\cal N}_{5}^2\right)-6 i a k_{x}^2 k_{y} {\cal N}_{3}  {\cal N}_{5} \Omega -2 i a k_{x}^3  {\cal N}_{4}  {\cal N}_{5}\Omega +6 i a k_{x} k_{y}^2  {\cal N}_{4}  {\cal N}_{5} \Omega +2 i a k_{y}^3   {\cal N}_{3}  {\cal N}_{5} \Omega \nonumber\\&-\gamma ^2 \Omega ^2+k_{z}^2 v_{z}^2 \Omega ^2+2 i \gamma  k_{z}v_{z} \Omega ^2- {\cal N}_{1}^2 \left( {\cal N}_{4}^2+ {\cal N}_{5}^2\right)- {\cal N}_{2}^2 \Big( {\cal N}_{3}^2+ {\cal N}_{5}^2\Big)- {\cal N}_{3}^2  {\cal N}_{5}^2- {\cal N}_{4}^2  {\cal N}_{5}^2\Big)^{1/2}.
\end{align}
$F(\beta,\gamma,\phi,\alpha)$ helps us to understand the nature of the eigen-energy. The expressions are a bit complicated. However, for all zero angles the $N_{i}$s could be simplified, and so is possible for $F(\beta,\gamma,\phi,\alpha).$ For non-zero $\beta,$ we discuss the band structure in the main text.  

\section{Light-driven single and triple WSM}
In this section, we would discuss the single and multi-WSM, illuminated by circular and BCL polarized light. 
\subsection{The single WSM with bi-circular light}
\begin{figure}\label{Fig6}
\includegraphics[width=0.4\textwidth]{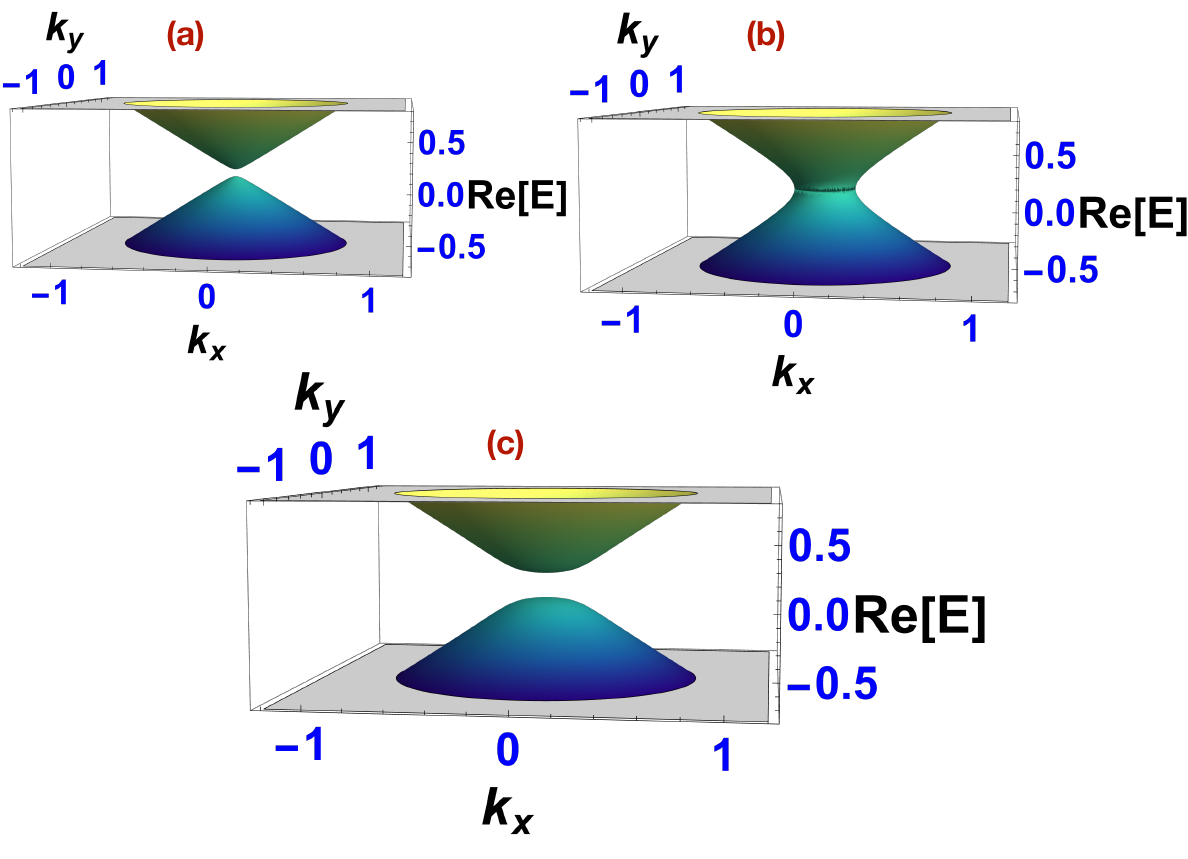}
\caption{${\cal E}_{\pm}$ vs $k_{x}$ and $k_{y}$ plots are presented. In (a), the NH parameter is considered to be zero. In the presence of the BCL light, we only have a gap opening in the system. In (b), at zero light amplitude, the NH parameter is switched on and the individual bands are broadened due to the presence of non-hermiticity. However, when both of the tuning parameters are present, the band gap still exists. Increasing $\gamma$ further does not contribute much in the spectra. The situation is depicted in Fig. (c). Thus, the role of BCL light is to diminish the effect of non-hermiticity in the presence of both parameters. However, in the absence of the BCL light, $\gamma$ changes the Weyl point to an exceptional contour.} 
\end{figure} 
In the main text we have elaborately discussed the triple WSM in the presence of the NH term and BCL polarized light. The natural question that can arise is whether the obtained results are specific to the triple Weyl or one can visualize similar effects with a normal (charge 1) WSM as well. To give a proper explanation, we add the results for the BCL light driven charge 1 NH WSM. The Hamiltonian of the system without driving is written as
\begin{align}
H(k)=\Big(v_{z}k_{z}+i\gamma\Big)\sigma_{z}+ a\Big(k^{}_{x}\sigma_{x}+k^{}_{y}\sigma_{y}\Big).
\end{align}
When driven by BCL light, the momentum $k_{x,y,z}$ are modified by the gauge potentials (Eq. (\ref{BCL})). Using the same technique as is in Eqs. (5) and (6) of the main text, we eventually have the effective Hamiltonian of the charge 1 WSM as
\begin{align}
H_{eff}=&\Big(v_{z}k_{z}+i\gamma\Big)\sigma_{z}+ a\Big(k^{}_{x}\sigma_{x}+k^{}_{y}\sigma_{y}\Big)+\frac{i}{\hbar \Omega}(\tilde{{\cal N}_{1}}\tilde{{\cal N}_{4}}-\tilde{{\cal N}_{2}}\tilde{{\cal N}_{3}})\sigma_{z}+\frac{i}{\hbar \Omega}(\tilde{{\cal N}_{1}}+\tilde{{\cal N}_{3}}){\tilde{\cal N}_{5}}\sigma_{y}\nonumber\\&-\frac{i}{\hbar \Omega}(\tilde{{\cal N}_{2}}+\tilde{{\cal N}_{4}})\tilde{{\cal N}_{5}}\sigma_{x},
\end{align}
where
\begin{align}
\tilde{{\cal N}_{5}}&=\frac{1}{2} i A_{0} v_{z}\sin (\phi ),~
\tilde{{\cal N}_{1}}=\frac{1}{2} A_{0} (a \cos (\beta )-i a \sin (\beta ) \cos (\phi ))\nonumber\\
 \tilde{{\cal N}_{2}}&=\frac{1}{2} A_{0} (a \sin (\beta )-i a \cos (\beta ) \cos (\phi )),~
  \tilde{{\cal N}_{3}}=\frac{1}{2} A_{0} (a \cos (\beta )+i a \sin (\beta ) \cos (\phi ))\nonumber\\
  \tilde{{\cal N}_{4}}&=\frac{1}{2} A_{0} (a \sin (\beta )+i a \cos (\beta ) \cos (\phi )).
\end{align}
The corresponding energy eigen values are 
\begin{align}
{\cal E}_{\pm}&=\pm\frac{1}{\Omega }\Big(a^2 k^{2}_{x} \Omega ^2+a^2 k^{2}_{y} \Omega ^2+2\tilde{{\cal N}_{2}} \left(\Omega  ((-i) a k_{x} \tilde{{\cal N}_{5}}-i k_{z} \tilde{{\cal N}_{3}} v_{z}+\gamma  \tilde{{\cal N}_{3}})+\tilde{{\cal N}_{1}} \tilde{{\cal N}_{3}}\tilde{{\cal N}_{4}}-\tilde{{\cal N}_{4}} \tilde{{\cal N}_{5}}^2\right)\nonumber\\&-2 i a k_{x}\tilde{{\cal N}_{4}} \tilde{{\cal N}_{5}} \Omega -2 \tilde{{\cal N}_{1}} \Big(-i a k_{y} \tilde{{\cal N}_{5}} \Omega +\tilde{{\cal N}_{4}} \Omega  (\gamma -ik_{z} v_{z})+\tilde{{\cal N}_{3}} \tilde{{\cal N}_{5}}^2\Big)+2 i a k_{y} \tilde{{\cal N}_{3}} \tilde{{\cal N}_{5}} \Omega -\gamma ^2 \Omega ^2+k_{z}^2 v_{z}^2 \Omega ^2\nonumber\\&+2 i \gamma  k_{z} v_{z} \Omega ^2+\tilde{{\cal N}_{1}}^2 \Big(-\Big(\tilde{{\cal N}_{4}}^2+\tilde{{\cal N}_{5}}^2\Big)\Big)-\tilde{{\cal N}_{2}}^2 \left(\tilde{{\cal N}_{3}}^2+\tilde{{\cal N}_{5}}^2\right)-\tilde{{\cal N}_{3}}^2 \tilde{{\cal N}_{5}}^2-\tilde{{\cal N}_{4}}^2 \tilde{{\cal N}_{5}}^2\Big)^{1/2}.
\end{align}
Then the corresponding energy (${\cal E}_{\pm}$) vs $k_{x}$ and $k_{y}$ plots are presented in Fig. 7. In Fig. 7(a), the NH parameter is considered to be zero. In the presence of the BCL light, we only have a gap opening in the system. In Fig. 7 (b), the NH parameter is switched on, and the individual bands are broadened due to the presence of non-hermiticity. However, the band gap still exists. Increasing $\gamma$ further does not contribute much in the spectra. Thus, the role of BCL light is to diminish the effect of non-hermiticity. However, in the absence of the BCL light, $\gamma$ changes the Weyl point to an exceptional contour. For charge 1 WSM, the system has no rotational symmetry. However the BCl has a rotational symmetry depending on the two frequencies of the constituents. This makes the charge 1 WSM gapped. In case of a triple WSM however, the rotational symmetry of the system is modified due to the rotational symmetry of the BCL. This makes it possible to achieve the unique Fermi surfaces.
\subsection{Triple WSM in presence of circularly polarized light}
	\begin{figure}\label{Fig7}
	\includegraphics[width=0.8\textwidth]{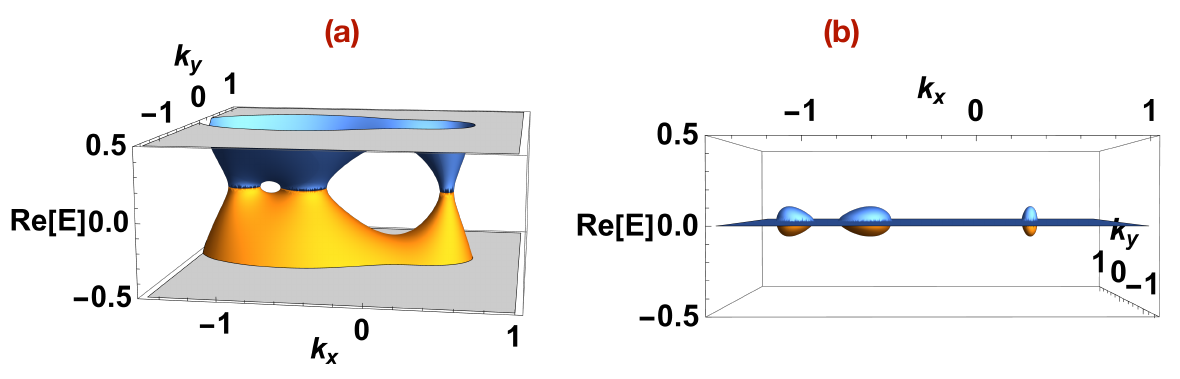}
	\caption{The circularly polarized light driven NH triple WSM eigen energy is presented for varying $k_{x}$ and $k_{y}.$ In (a) the real part of the eigenenergy is plotted for $A_{0}=0.8$ and $\gamma=0.3.$ Here one can notice the three exceptional contours appear due to rotational symmetry-breaking terms. In (b), the imaginary eigen-energy is presented. } 
	\end{figure} 
On a similar note, one may wonder about the effects one visualises when a simple circular polarized light is illuminated on the triple WSMs. Although this effect is briefly discussed in our previous paper \cite{Our}, for the completeness of this appendix, we have elaborated the band structure here. The starting Hamiltonian is the same as Eq. (\ref{triple}).  
Next, let us illuminate the triple weyl with a circularly polarized light. This modifies the momentum as 
\begin{align}
k_x\rightarrow k_x+A_{0} \sin(\Omega t)\nonumber\\
k_z\rightarrow k_y+A_{0} \cos(\Omega t),	
\end{align}
where $A_{0},$ and $\Omega$ are the driving field amplitude and frequency respectively. Using the high frequency approximation we eventually have the effective Hamiltonian as follows
	\begin{align}\label{A7}
	&H_{\mathrm{eff}}(k)=a\Big((k_{-}^{3}+i \Delta_{}k_{-}^{2}) \sigma_{+}+(k_{+}^{3}-i \Delta_{}k_{+}^{2})\sigma_{-}\Big)+\left(v_{z} k_{z}+i\gamma\right)\sigma _z,
	\end{align}
		where $\Delta_{} =\frac{3A_{0}^{2}v_{z}}{2\hbar \Omega}.$
Also $k_{\pm}=k_{x}\pm i k_{y}$ and $\sigma_{\pm}=\sigma_{x}\pm i \sigma_{y}$ and $\gamma$ is the gain/loss parameter. The corresponding energy eigenvalues are written as
		\begin{align}\label{13}
	&{\cal E}^{}_{\pm}=\pm\sqrt{\frac{(k_{x}^{2}+  k_{y}^{2})^{2} (k_{x}^{2} + (k_{y} - \Delta_{})^2) }{m^{2}}-(\gamma - 
		i k_{z} v_{z} \eta)^2}.
	\end{align}
	Importantly, in this case, the charge $3$ WSM, goes through a rotational symmetry breaking, and eventually, we obtain three charge 1 exceptional contours as is obtained in FIG. 8. Fig. 8 (a) and (b) show the real and imaginary parts of the energy spectra. 

\end{widetext}

\end{document}